\documentclass[12pt,preprint]{aastex}
\begin{document}
\title{Hubble Space Telescope STIS Spectroscopy 
of Three Peculiar Nova-Like Variables:
BK Lyn, V751 Cygni and V380 Oph}

\author{Robert Zellem, Nicholas Hollon, Ronald-Louis Ballouz, 
Edward M. Sion, Patrick Godon\altaffilmark{1}}

\affil{Dept of Astronomy \& Astrophysics,
Villanova University,
Villanova, PA 19085: robert.zellem@villanova.edu, 
nicholas.hollon@villanova.edu, 
ronald-louis.ballouz@villanova.edu, edward.sion@villanova.edu, 
patrick.godon@villanova.edu}
                           
\author{Boris T. G\"ansicke}
\affil{Department of Physics, University of Warwick, 
Coventry, CV4 9BU, UK: Boris.Gaensicke@warwick.ac.uk}

\author{Knox Long}
\affil{Space Telescope Science Institute, Baltimore, MD 21218: 
Long@stsci.edu}
\altaffiltext{1}
{Visiting at the Space Telescope Science Institute, Baltimore,
MD 21208; godon@stsci.edu}

\begin{abstract}

We obtained Hubble STIS spectra of 
three nova-like variables: V751 Cygni, V380 Oph,  
and - the only confirmed nova-like variable known to be below the 
period gap - BK Lyn. 
In all three systems, the spectra were taken during high optical 
brightness state, and a luminous accretion disk
dominates their far ultraviolet (FUV) light. 
We assessed a lower limit of the distances by 
applying the infrared photometric method of \citet{Knigge2006}. 
Within the limitations imposed by the poorly known system
parameters (such as the inclination, white dwarf mass, and the 
applicability of steady state accretion disks) 
we obtained satisfactory fits to BK Lyn using optically thick 
accretion disk models with 
an accretion rate of 
$\dot{M} = 1\times10^{-9} M_{\odot}$ yr$^{-1}$ for a white dwarf mass of 
$M_{wd} = 1.2 M_{\odot}$ and 
$\dot{M} = 1 \times 10^{-8} M_{\odot}$ yr$^{-1}$ for $M_{wd} = 0.4 M_{\odot}$.
However, for the VY Scl-type nova-like variable V751 Cygni
and for the SW Sex star V380 Oph, we are unable to obtain 
satisfactory synthetic spectral fits to the high state FUV spectra 
using optically thick steady state accretion disk models.  
The lack of FUV spectra information down to the Lyman limit hinders 
the extraction of information about the accreting white dwarf
during the high states of these nova-like systems. 
\end{abstract}


\section{Introduction}

Cataclysmic variables (CVs) are short-period binaries 
in which a late-type, Roche-lobe-filling main-sequence 
dwarf transfers gas through an accretion disk
onto a rotating, accretion-heated white dwarf (WD).
The nova-like variables are a subclass of CVs in which
the mass-transfer rate is high and the light of the
system is dominated by a very bright accretion disk.
The spectra of nova-like variables resemble those of
classical novae (CN) that have settled back to quiescence. 
However, nova-like variables have never had
a recorded CN outburst. Hence their evolutionary
status is unknown. They could be close to having
their next CN explosion, or they may have had an unrecorded 
explosion, possibly hundreds or thousands
of years ago. Adding to the mystery of nova-like variables 
is that some of them (known as the VY Sculptoris stars after their prototype) show the curious 
behavior of being in a high optical brightness state most
of the time, but then, plummeting into a deep low-brightness state with little or
no ongoing accretion. Then, just as unpredictably,
they return to the high-brightness state. These drops
are possibly related to cessation of mass transfer from
the K-M dwarf secondary star either by starspots positioned under L1 
\citep{Livio1994}        or 
irradiation feedback in which an inflated outer disk can modulate the mass transfer
from the secondary by blocking irradiation of the hot
inner accretion disk region \citep{wu1995}.   

As part of a Hubble Space Telescope Snapshot project 
extending over three Hubble observing cycles (B. G\"ansicke, PI),
several nova-like variables were observed
among dozens of non-magnetic and magnetic cataclysmic variables.
In this paper we report on an analysis of the HST STIS spectra
of three of these objects, BK Lyn, V751 Cygni and V380 Oph. 

\subsection{BK Lyn}

BK Lyn (PG0917+342) is an extraordinary nova-like variable with an orbital period placing it below
the CV period gap. As such, 
it is the only bona fide nova-like variable below the CV period
gap \citep{Dobrzycka1992}. It was classified a short period 
dwarf nova by 
\citet{Dhillon2000} but long term light curves from 
ROBOSCOPE (98 exposures over
three years) and the Harvard Plate Collection
reveal no evidence for dwarf nova outbursts or even low optical brightness states.
The system normally remains between V = 14.6 and 14.7.
\citet{Ringwald1996} 
confirmed that its orbital period places it below the period gap.
\citet{Skillman1993} detected
permanent superhumps and found that the optical spectrum fits a power law 
$f_{\lambda} = k_{\lambda}^{-\alpha}$ of index $\alpha$ = 2.66 $\pm$ 0.10 characteristic of a steady state 
accretion disk. \citet{Dhillon2000} 
carried out infrared spectrophotometry of BK Lyn
revealing a secondary spectral type of M5V with 
an upper limit to the contribution
of the secondary to the K-band light of 50\% $\pm$ 5\%. 
\citet{Dobrzycka1992} estimated the binary's inclination to be 
32 $\pm$ 12 degrees, white dwarf mass of 0.3 $^{+0.35}_{-0.12}$. 
However, the white dwarf is not well-constrained. 
\citet{Sproats1996} give a distance estimate of 114 pc 
using K magnitudes and Bailey's method, while 
\citet{Dhillon2000} give a lower limit of 185 pc using IR spectroscopy. 
The distance to the system is crucial for this analysis.
\citet{Puebla2007} cite a distance of 150 pc which appears to 
be the average of the \citet{Sproats1996} distance and the \citet{Dhillon2000}
lower limit of 185 pc. \citet{Dobrzycka1992}
gave $\sim$ 1000 pc but based it on assuming the average absolute visual magnitude for nova-like variables of +4.0.
We have used the method of \citet{Knigge2006} 
to obtain a strict lower limit distance of 116 pc.

\subsection{V751 Cygni}

V751 Cygni is classified as a VY Scl system.  A far ultraviolet IUE
spectrum of V751 Cygni was taken on April 25, 1985 when the system was at
or near optical maximum. The spectrum, SWP25574, was obtained at low
dispersion through the large aperture. The visual magnitude with the fine
error sensor (FES) was V(FES) = 14.2 while a longer wavelength UV
spectrum, LWP05819, yielded an interstellar reddening value E($\bv$) =
0.25$\pm$0.05 from the 2200A feature \citep{Greiner1999}. 
The large amount of reddening introduces additional uncertainty.
Based upon the
lack of emission lines and upon a study by \citet{Greiner1998} of equivalent
width versus orbital inclinations for VY Scl stars, \citet{Greiner1999}
concluded that a low inclination ($< $50$\degr$) is indicated. They
estimate from the \citet{Greiner1998} study that $i$ = 30$\pm$20 $\degr$. 
\citet{Greiner1999} do mention a possible He II (1640) absorption feature. In
fact, this feature, if real, is the strongest absorption line in the IUE
spectrum. However, while \citet{Greiner1999} measured equivalent widths
for the SiV, C IV and N V, which are only marginally present in the IUE
spectrum, they do not give further mention to the strongest absorption
feature near 1640A. 

The donor star was classified early to mid-F star, which was implied by
the presence of the G-band and Na D lines with absence of Mg b 5169A
\citep{Downes1995}. However, subsequent work demonstrated
that the feature thought to be the G-band was actually part of the
H$\gamma$ wing while the Na D lines proved to be interstellar. There are
also reports of transient optical P Cygni absorption which appears
correlated with binary phase \citep{Patterson2001}.  The
mass of the white dwarf is unknown. The radial velocity studies done to
date using disk emission lines, which may reflect the dynamical motion of
the white dwarf, have been inconclusive 
\citep{Walker1980,Echevarria2002,Downes1995} ). 

\citet{Greiner1999} carried out a photometric and spectroscopic analysis
of the system and reported negative superhumps and the unexpected detection of soft X-rays instead
of the predicted hard X-rays, during the low optical brightness state of
V751 Cygni when the boundary layer of the WD is optically thin. They
suggest that V751 Cyg is a supersoft X-ray binary, with a white dwarf
undergoing mass accretion at a high enough rate to cause quasi-steady
hydrogen burning on the surface. This would imply the white dwarf is
highly luminous (L$_{bol}$ $\sim$ 10$^{36}$ - 10$^{38}$ ergs s$^{-1}$).
\citet{Patterson2001} confirm the soft X-ray detection during quiescence
but contend that V751 Cyg is not a supersoft binary, but has a low
luminosity soft X-ray component ($\sim$ 10$^{33}$ erg s$^{-1}$). Clearly,
the mass of the white dwarf in this systems is critical.  
V751 Cyg could also be an SW Sex star as well. It has the right orbital period as well as
negative superhumps in the system. It is important
therefore to explore the physical properties of the disk and, if possible, underlying, accreting
white dwarf in V751 Cygni.

\subsection{V380 Oph}

The third system, V380 Oph, is classified as an SW Sextantis system, another nova-like subclass,
characterized by a multitude of observational characteristics: orbital periods between 3 and 4 hours, 
one third of the systems non-eclipsing and two-thirds showing deep eclipses of the WD by the 
secondary, thus requiring high inclination angles, single-peaked emission lines despite the high inclination, 
and high excitation spectral features including He II (4686) emission and strong Balmer emission on a 
blue continuum, high velocity emission S-waves with maximum blueshift near phase $\sim$ 0.5, delay of emission line
radial velocities relative to the motion of the WD, and cental absorption dips in the emission lines
around phase $\sim$ 0.4 - 0.7 
\citet{Rodriguez-Gil2007,Hoard2003}.
The white dwarfs in many, if not all, of these systems are suspected of 
being magnetic \citep{Rodriguez-Gil2007}. 
Since these objects are found near the 
upper boundary of the  period gap 
(\citet{Warner1995} and references therein), 
their study is of critical importance to understanding
CV evolution as they enter the period gap. 

\citet{Shafter1983} determined a spectroscopic period for V380 
Oph from observations with very moderate spectral resolution. 
\citet{Rodriguez-Gil2007} improved its orbital period and
measured H-alpha emission radial velocity variations with an amplitude of
~400 km/sec. Photometric observations have revealed orbital 
variability with possible negative superhumps.

\subsection{Reddening Values and Distances} 

The reddening of the systems was taken 
from available estimates in the literature. 
The three principal sources of reddening for
cataclysmic variables are the compilations of 
\citet{Verbunt1987,LaDous1991,Bruch1994}. For V751 Cygni, \citet{Greiner1999}
gave E($\bv$) = 0.25$\pm$0.05 which is the value we adopted for V751 Cygni.
For V380 Oph, we treated the reddening as a free parameter in the model fitting.
The FUV spectra of V751 Cygni objects were de-reddened using the IUERDAF 
IDL routine UNRED. The spectrum of BK Lyn was not dereddened as the
galactic reddening in the direction of BK Lyn is very small
$(E(B-V) \sim 0.01$. 
The observed properties of all three systems are summarized in Table 1.

We have computed strict lower limit distances to BK Lyn, V751 Cygni and V380 Oph using a 
method by \citet{Knigge2006} 
which uses 2MASS JHK photometry. For each system, 
we obtained the J,H,K apparent magnitudes from 2MASS. For a given orbital period, \citet{Knigge2006}
provides absolute J, H and K magnitudes based upon his semi-empirical donor sequence for 
CVs. If it is assumed that the donor provides 100\% of the light in J, H and K, then the 
distance is a strict lower limit. If the donor emits 33\% of the light (the remainder being
accretion light), then a very approximate upper limit is obtained. At the K-band, the latter limit is a factor of
1.75 times the lower limit distance. The lower limit distances are used as constraints in the 
synthetic spectral fitting procedures described below.
For BK Lyn, this leads to a lower limit distance of 116 pc, for V380 Oph, this method yields a range of distance from 293 pc (strict lower limit) to 
513 pc while for V751 Cygni, we obtained a range 201 pc (strict lower limit) to 352 pc.

\section{Observations and Data Reduction}

FUV spectroscopy of BK Lyn, V751 Cygni and V380 Oph was obtained with {\it{HST}}/STIS
during {\it{HST}} Cycle 11. The data were obtained using the G140L grating
and the $52^{\prime\prime} \times 0.2^{\prime\prime}$ aperture, providing a
spectral resolution of R$\sim 1000$ over the wavelength range 1140-1720
\AA. Since the total time involved in each snapshot observation was short
($< 35$min), the observations were made in the ACCUM mode in order to
minimize the instrument overheads. This resulted in exposure times of 600
to 900 seconds.  Each snapshot observation resulted in a single time
averaged spectrum of each system. All of the spectral data were
processed with IRAF using CALSTIS V2.13b.During target acquisition,
{\it{HST}} points at the nominal target coordinates and takes a
$5^{\prime\prime} \times 5^{\prime\prime}$ CCD image with an exposure time of a few seconds.
Subsequently, a small slew is performed that centers the target in the
acquisition box, and a second CCD image is taken. The acquisition imagesfor these observations were obtained using the F28x50LP long-pass
filter, which has a central wavelength of 7228.5 and a full-width at half
maximum (FWHM) of 2721.6 \AA 
\citep{Araujo-Betancor2005}. 

The instrumental setup and exposure details of the {\it{HST}} STIS spectra
of BK Lyn, V380 Oph and V751 Cygni are provided in the
observing log given in Table 2, the entries are by column: (1)  the
target, (2) Data ID, (3) Instrument Config/Mode, (4) Grating, (5)
Aperture, (6) Date of Observation (yyyy-mm-dd), (7) Time of observation,
and (8) Exposure time (s).

\subsection{BK Lyn}

Until 2003, there were no FUV spectra of BK Lyn with which to check for P Cygni profiles 
indicating wind outflow or an analyze the FUV spectral slope or FUV line profiles.
As part of an HST snapshot program (see above), the first FUV spectrum of BK Lyn 
was secured. The spectrum is displayed in Fig.1 where a steeply rising continuum
toward shorter wavelengths is seen together with strong emission 
features at C III (1175), NV (1240), Si III + OI (1300), C II (1335),
Si IV (1400), C IV (1550) and weak He II (1640) emission. 
The emission lines suggest it is probable that an accretion disk is present in the system at the time 
of our HST spectrum. The continuum
flux level ranges from $\sim$ 2 x 10$^{-14}$ ergs/cm$^{2}$/s/\AA\ 
at the short end to
$\sim$ 1.2 $\times$ 10$^{-14}$ ergs/cm$^{2}$/s/\AA\ 
at the long wavelength end. The presence of 
strong emission lines coupled with the low flux level and the steeply
rising continuum are not characteristic of the FUV spectra of nova-like variables
in their high optical brightness states unless the inclination angle is high.
We note that the emission lines and continuum slope in BK Lyn's spectrum are strikingly similar in appearence 
to the HST and HUT spectra of the dwarf nova SS Cygni in quiescence (see Fig.4 in Long et al.2005) as well as
the FUV spectrum of V794 Aql in its high state (Godon et al.2007). 
The F28$\times$50LP magnitude from the acquisition exposure was 15.3, roughly consistent
with a high state but slightly fainter than the normal visual magnitude range of 14.6 to 14.7.

\subsection{V751 Cygni}

The STIS spectrum of V751 Cygni reveals moderately strong absorption features seen against a continuum rising toward
shorter wavelengths. The FUV continuum even after de-reddening, has a relatively flat slope while the Ly$\alpha$
profile is rather narrow. This poses a considerable challenge in finding an accretion disk spectral energy distribution
which satisfies both the continuum slope and yields the observed Ly$\alpha$ line width in a single disk model.  
The narrow width of Ly$\alpha$ should be associated with a much steeper 
continuum slope than is observed in the spectra. Our HST STIS spectrum confirms the identification of
He II (1640) in absorption although the equivalent width of the feature is
smaller than seen in the IUE spectrum (see Section 1.1 above). Thed C IV (1550) absorption lines reveals a hint of P Cygni structure
probably indicating wind noutflow at the time the STIS spectrum was obtained.

\subsection{V380 Oph}

The STIS spectrum of V380 Oph is dominated by strong broad emission features superimposed on a continuum rising 
toward shorter wavelengths For V380 Oph. We found the same inconsistency between the observed width  of the Ly$\alpha$ profile and 
the observed continuuum slope. Several weak, sharp absorption features are most likely of interstellar origin. 
For V380 Oph, we found the same inconsistency between the observed width  of the Ly$\alpha$ profile and the observed 
continuuum slope. This led us to first explore the effect of interstellar reddening on the spectrum.

In table 3, we have listed the strongest spectral features detected for each system.

\section{Multi-Component Synthetic Spectral Models}

We adopted model accretion disks from the optically thick disk model grid 
of \citet{Wade1998}. In these accretion disk models, the innermost 
disk radius, 
R$_{in}$, is fixed at a fractional white dwarf radius of $x = R_{in}/R_{wd} = 1.05$. 
The outermost disk radius, R$_{out}$, was chosen so that T$_{eff}(R_{out})$ is near 10,000K
since disk annuli beyond this point, which are cooler zones with larger radii, would provide
only a very small contribution to the mid and far UV disk flux, particularly the SWP FUV bandpass. 
The mass transfer rate is assumed to be the same for all radii.
Thus, the run of disk temperature with radius is taken to be:

\begin{equation}
T_{eff}(r)= T_{s}x^{-3/4} (1 - x^{-1/2})^{1/4}
\end{equation}

where  $x = r/R_{wd}$
and $\sigma T_{s}^{4} =  3 G M_{wd}\dot{M}/8\pi R_{wd}^{3}$

The disk is divided into a set of ring annuli. The vertical structure of each
ring is computed with TLUSDISK \citep{Hubeny1990}, 
which is a derivative of the stellar atmosphere
program TLUSTY \citep{Hubeny1988}. The spectrum synthesis program SYNSPEC
described by \citet{Hubeny1995} is used to solve
the radiative transfer equation to compute the local, rest
frame spectrum for each ring of the disk. In addition to
detailed profiles of the H and He lines, the spectrum synthesis
includes metal lines up to Nickel (Z = 28). The accretion disks are computed in LTE and
the chemical composition of the accretion disk is kept fixed at solar
values in our study.

Limb darkening of the disk is fully taken into account in the 
manner described by \citet{Diaz1996} involving
the Eddington-Barbier relation, the increase of kinetic temperature
with depth in the disk, and the wavelength and temperature dependence of the Planck function.

The details of our $\chi^{2}_{\nu}$ (per degree of freedom) minimization fitting procedures are 
discussed in detail in \citet{Sion1995}.
We take any reliable published parameters 
like the WD mass, or orbital inclination only as an initial guess in searching for the best-fitting accretion disk models. 
We relax all constraints except the lower limit \citet{Knigge2006} distance. For each systems's spectrum, 
we carry out fits for every combination of 
$\dot{M}$, inclination and white dwarf mass
in the \citet{Wade1998} library. 
The values of {\it{i}} are 18, 41, 60, 75 and $81\degr$. 
The range of accretion rates covers $-10.5 < \log \dot{M} < -8.0$ in steps of 0.5 in the log and five different 
values of the white dwarf mass, namely, 0.4, 0.55, 0.8, 1.0, and 1.2 M$_{\sun}$. 
Out of the roughly 900 models using every combination of $i$, \.{M} and M$_{wd}$, we try to isolate
the best-fitting accretion disk model.  In effect, we search for the best fit
based upon (1) the minimum $\chi^{2}$ value achieved, (2) the goodness of
fit of the continuum slope, (3) the goodness of fit to the observed Ly$\alpha$ 
region and (4) consistency of the scale factor-derived distance with
the adopted distance or distance constraint. In other words, the fitting solution may not necessarily be the model 
with the lowest $\chi^{2}$ value but rather all other 
available constraints are used such as constraints on the distance, reliable system parameters
if available, absorption line fits, especially the Ly$\alpha$ wings but
sometimes even Si II, Si III, C III, C II, Si IV and Si II when these
features are in absorption and do not have an origin in a wind or
corona. We also search for any statistically significant improvement in the
fitting by combining white dwarf models and disk models together using a
$\chi^{2}$ minimization routine called DISKFIT.  Once again, we find
the minimum $\chi^{2}$ value achieved for the combined models, and check
the combined model consistency with the observed continuum slope and Ly$\alpha$ 
region, and consistency of the scale factor-derived distance with
the adopted distance or distance constraints. For the white dwarf radii,
we use the mass-radius relation from the evolutionary model grid of 
\citet{Wood1990} for C-O cores.

\section{Theoretical Model Fitting Result}

For a non-magnetic nova-like variable during its high state, 
it is reasonable to expect
that a steady state optically thick accretion disk 
might provide a successful fit (see \citet{Hamilton2008}).
This was our intial expectation for all three nova-likes, BK Lyn, V380 Oph and V751 Cygni.

For BK Lyn, the most striking feature of its spectrum is the sharp, narrow Ly$\alpha$ line in combination with 
a steeply rising continuum. The narrowness of the Ly$\alpha$ and steep continuum slope suggested the possiblity
that the FUV flux originates from a hot white dwarf photosphere. But a fit to the the continuum slope with a hot photosphere
at T=30000K and a nominal white dwarf gravity of log g=8.0, yields a synthetic Ly$\alpha$ absorption feature which is far too broad.
However, a narrower profile is obtained by increasing the effective temperature and/or lowering the gravity. 
The continuum slope and narrow Ly$\alpha$ feature of the observed spectrum can be matched with a white dwarf photosphere
with T=38000K and log g=7. A surface gravity this low would not be inconsistent
with the low white dwarf mass derived by 
\citet{Dobrzycka1992}. However, the distance implied by the fit is
1400 pc while the stellar radius approached that of a hot subdwarf. Moreover, the V-band magnitude of this model fit is 20.6, 
which is about five magnitudes fainter than the observed magnitude. 

The narrowness of the Ly$\alpha$ profile also suggested the possibility that it is of interstellar origin.
First, if the system were being viewed at low inclination, then the emission lines which are presumed to be formed 
in the disk, would not be as broad as observed. This makes it likely that the narrow Ly$\alpha$ absorption, given that
it is not due to a low gravity white dwarf (see above), is due to interstellar atomic hydrogen. The reddening toward BK Lyn is 
rather small, possibly as small as $\sim 0.01$ but the corresponding density is enough to produce such an absorption. 
One needs an H column density of only $\sim 1 \times 10^{19}$cm$^{-2}$, which
is a small fraction of the galactic column in line of sight to BK Lyn 
which is $\sim 1.2 \times 10^{20}$cm$^{-2}$.  

Since a white dwarf photosphere is unlikely to provide the FUV flux, we tried fitting accretion disk models alone to the STIS data. 
Because the narrow Ly$\alpha$ line is almost certainly interstellar, the mass of the white dwarf, inclination angle and distance are
essentially unconstrained except for the 
lower limit to the distance given by the 
\citet{Knigge2006} method. Moreover, the presence of 
broad emission lines and the optical radial velocity amplitude of +/- 100 km/s 
rules out a face-on or very low inclination system. Applying our fitting procedures described above, we found accretion disk model 
yielded much better fits in every respect than a white dwarf photosphere. We also tried white dwarf plus accretion disk combination
fits but we found that the best fits resulted only when the white dwarf was contributing 1\% or less of the FUV flux.
Thus our best fitting solutions result when the white dwarf flux contribution is negligible.
In Table 4 we have summarized the disk model fitting results to the spectrum of BK Lyn.

In Figures 1 ,2 and 3, we display the best-fitting accretion disk models for
a 1.2 M$_{\sun}$ WD, a 0.8 M$_{\sun}$ WD and a 0.4 M$_{\sun}$ WD. we find that the disk fitting
favors high inclination with the accretion rate of BK Lyn lying between 
$\dot{M} = 1\times10^{-8} M_{\sun}$ yr$^{-1}$
and 
$\dot{M} = 1\times10^{-9} M_{\sun}$ yr$^{-1}$ for the highest 
($1.2 M_{\sun}$) and lowest ($0.4 M_{\sun}$) white dwarf mass respectively.
It is likely that at this inclination, the White dwarf 
could be masked by the swollen disk. 

Applying the same fitting procedures to V751 Cygni, We found no combination of parameters in the accretion disk model grid which provided any 
acceptable agreement with the de-reddened (E(B-V) = 0.25) STIS spectrum of V751 Cygni. The only solution that even remotely resembled the STIS spectrum (but still in very poor agreement),
is an accretion disk model with a white dwarf mass, 
M$_{wd} = 0.55$ M$_{\sun}$, orbital inclination ${i}  = 75\degr$, a distance of only 78 pc, with an accretion rate of 
$\dot{M} = 1\times10^{-9} M_{\sun}$ yr$^{-1}$. The $\chi^{2}_{\nu}$  value for this fit is $\chi^{2}_{\nu}  = 11.94$.
Increasing the dereddening value up to 0.5 decreases the distance even more. Thus we are forced to conclude that there is no optically thick steady state disk model that fits the FUV 
spectrum of V751 Cygni. In Fig. 4, we display the STIS spectrum of V751 Cygni together with an accretion disk model that illustrates the extent of the disagreement 
between standard disk models and the observed data.
   
Unfortunately, the situation is only slightly better for V380 Oph. 
We encountered the same difficulty finding acceptable fits. 
As stated earlier, this system also has the same 
inconsistency between the observed Ly$\alpha$ profile width 
and the observed continuuum slope as we found for V751 Cyg. This led us to 
first explore the effect of interstellar reddening on V380 Oph's spectrum.
In our disk model fitting therefore we treated E(B-V) as a free parameter. 
We found that
for a color excess E($\bv$) = 0.20, the observed spectral slope and 
observed Ly$\alpha$ improved slightly but considerable inconsistency 
remained. 
As in the case of V751 Cygni, we carried out disk model 
fits to V380 Oph for 
every combination of \.{M}, inclination and white dwarf mass in our disk model library.  
The only model fit to the STIS data that 
even slightly resembled (but still in poor agreement with) 
the STIS spectrum was an accretion disk model alone. 
For E($\bv$) = 0.20, this closest but still rather poor fit corresponded to
M$_{wd} = 0.4$ M$_{\sun}$, $\it{i}$ = 75$\degr$ 
with $\dot{M}$ = 1$\times$10$^{-8}$ $M_{\sun}$ yr$^{-1}$ 
for a $\chi^{2}_{\nu}$  = 13.76 and a distance of 312 pc. 
In Fig.5, we illustrate the extent of the disagreement 
between the STIS spectrum of V380 Oph and a standard accretion disk model. 
The fitting of the Ly$\alpha$ absorption feature has been achieved with
the inclusion of the ISM model. The ISM model has an hydrogen column
density of $3 \times 10^{20}$cm$^{-2}$, a turbulent velocity of 
50km/s and a temperature of T=170K. Without the ISM model (dashed line)
the disk model does not fit the bottom of the Ly$\alpha$ profile.  

It is clear that for V380 Oph, a definite SW Sex star, and for V751 Cygni (itself is a possible SW Sex member),
optically thick steady state accretion disk models simply cannot be fit successfully to their FUV spectra. 
If one examines their FUV to optical spectral energy distribution, both V380 Oph and V751 Cygni look too red
for a standard accretion disk SED in terms of UV-optical colour.
On the other hand, such a conclusion may not apply to all definite SW Sex stars. For example, we point out here that
we have found that the FUV spectra of of two definite SW Sex systems V442 Oph and AH Men in their high state do reveal
reasonable agreement with steady state, optically thick disk models. In Figures 6 and 7, we display accretion disk fits to 
V442 Oph and AH Men from \citet{Ballouz2009} to illustrate this point. 
We note that while the disk fits are not perfect to these two SW Sex stars, 
the results do suggest that for some SW Sex stars, optically thick, 
steady state disk models do provide a fair representation of the FUV energy distribution and 
Ly$\alpha$ regions. 

\section{Summary}

(1) We obtained and analyzed the first FUV spectrum of BK Lyn, the only confirmed nova-like variable known to be
below the CV period gap. The spectrum is characterized by a steeply rising continuum
toward shorter wavelengths together with strong broad emission 
features at C III (1175), NV (1240), Si III + OI (1300), C II (1335),
Si IV (1400), C IV (1550) and weak He II (1640) emission. As noted earlier, the emission lines and continuum slope in BK Lyn's spectrum
are similar in appearence to the HST and HUT spectra of the dwarf nova SS Cygni 
and to the FUV spectrum of V794 Aql in its high state.
The emission lines are presumed to have formed in an accretion disk present in the system at the time 
of our HST spectrum. A very narrow Ly$\alpha$ absorption feature is seen which is almost certainly
due to interstellar hydrogen. No direct evidence of wind outflow in the form of P Cygni-type profiles is seen
in the HST spectrum. We find that BK Lyn has an accretion rate in the 
range $\dot{M} = 1\times10^{-9} M_{\sun}$yr$^{-1}$ 
if $M_{wd} = 1.2 M_{\sun}$ 
and 
$\dot{M} = 1\times10^{-8} M_{\sun}$yr$^{-1}$ if $M_{wd} = 0.4 M_{\sun}$. 

If mass transfer in CV
systems below the period gap is driven by angular momentum loss due to
gravitational wave emission alone , then the predicted accretion rate at
an orbital period of 108 minutes is 5 $\times$ 10$^{-11}$ M$_{\sun}$
yr$^{-1}$ (Howell, Nelson \& Rappaport 2001), a value much lower than 
the likely average accretion rate of a nova-like system like BK Lyn. 

\citet{Puebla2007} noted the peculiarity of 
BK Lyn and stated that there was no way to 
fit the {\it HST}/STIS spectrum of BK Lyn with 
their accretion disk model. They pointed out that 
it is possible BK Lyn has an optically thin disk with a very 
low accretion rate. However, they adopted a distance of  
only d=150pc (presumably the average of Sproat's distance 
of 114 pc and Dhillon's distance of  
185pc) they determined an accretion rate of 
$\dot{M} = 3.3\times10^{-10} M_{\sun}$yr$^{-1}$. Taking their 
low accretion rate,  BK Lyn should be a 
dwarf nova. However, an outburst has never 
been observed either by Roboscope, Rotse or by the
AAVSO and Vsnet.

Nova-like variables are generally associated with higher
average mass transfer rates than dwarf novae. Indeed, an
examination of the white dwarf temperatures in CVs below the period gap
are strongly clustered around 15,000K (Szkody et al.2003, Gaensicke et al.
2004) with the hottest example being VW Hyi with T$_{eff}$ = 20,000K. 
Since there is as yet no evidence of low brightness states in BK Lyn, the accretion rate may remain
high for long intervals of time. Therefore, in view of the higher average mass transfer rate compared with dwarf novae
below the gap, it would be most interesting to know if the underlying white dwarf's surface
temperature, which is controlled by long term compressional heating, is hotter than the nominal 15,000K  which is typical of white dwarfs in
CVs below the gap.
 
For V751 Cygni, a VY Scl-type nova-like, we carried out an analysis of the de-reddened HST STIS spectrum and an 
archival IUE spectrum obtained near maximum brightness, using the lower limit distance obtained by the \citet{Knigge2006} 
method to help constrain the fits. The fact that the FUV spectrum 
is dominated by absorption features suggests that the inclination may be low.
If, according to \citet{Greiner1999}, it is i = 30 +/- 20 degrees, then 
it is very likely that 
one is observing at least some fraction of the upper hemisphere of the white dwarf as well as the dominant luminous accretion
disk. Therefore, it would most interesting to gain an estimate of the white dwarf temperature.
Unfortunately, during the high state of a nova-like variable viewed at low inclination, the accretion disk light normally dominates and it is difficult
to constrain the temperature of the accreting white dwarf without spectra such as those obtained
with FUSE, which extend down to the Lyman limit where in some cases a disk and photosphere present distinguishable signatures. 
The slope of the continuum in V751 Cygni is relatively flat even after the large de-reddening while the Ly$\alpha$ absorption is narrow.
We were unable to find any optically thick, steady state accretion disk model that woiuld yield a satifactory fit to the STIS spectrum of V751 Cygni.
Therefore, we could not determine a reliable accretion rate for this system. As stated earlier, V751 Cygni could turn out to be an SW Sextantis subclass of nova-like variables.
Many such systems reveal reddish UV-optical spectral energy distributions and thus cannot be fit with standard disk models.

For V380 Oph, a definite SW Sextantis nova-like system, the FUV spectrum is dominated by broad emission features, a relatively flat continuum and a narrow Ly$\alpha$ absorption feature.
Even with a reddening of E(B-V)= 0.2 and higher, our synthetic spectral analysis using standard accretion disk models was unable to derive a reliable accretion 
rate for V380 Oph. This is another example of an SW Sex star with a reddish UV-optical spectral energy distribution that cannot be fit successfully with standard disk models.
On the other hand, as shown in Figures 6 and 7, there are some definite SW Sex systems can be fit sucessfully with standard disk models. Like V751 Cygni, we can say little about the white dwarf in V380 Oph during it
high state without FUV spectra which extend down to the Lyman Limit.
Further determinations of accretion rates and, whenever possible, white dwarf temperatures
are required to elucidate the role of SW Sex stars and other nova-like variables in 
CV evolution.

\acknowledgements

This work was supported by HST grant snapshot grants GO-09357.02A, 
GO-9724.02A NSF grant AST0807892, NASA grant NNG04GE78G and by 
summer undergraduate research support from the NASA-Delaware 
Space Grant Consortium. The ISM model used in this work was generated by 
Paul E. Barrett (USNO) for the analysis of the {\it FUSE} spectra of
DNs and NLs related to a different project (PI Godon).  
PG wishes to thank Mario Livio for his kind hospitality as the
Space Telescope Science Institute, where part of this work was
carried out. Some or all of the data presented in this paper were obtained from 
the Multimission Archive at the Space Telescope Science Institute (MAST). 
STScI is operated by the Association of Universities for Research 
in Astronomy, Inc., under NASA contract NAS5-26555. Support for 
MAST for non-HST data is provided by the NASA Office of Space 
Science via grant NAG5-7584 and by other grants and contracts.

\clearpage

\begin{deluxetable}{lccccccccc}
\tablecaption{System Parameters}
\tablecolumns{9}
\tablenum{1}
\tablehead{
\colhead{System}
&\colhead{V$_{max}$}
&\colhead{V$_{min}$}
&\colhead{P$_{orb}$ (d)}
&\colhead{{\it{i}} ($\degr$)}     
&\colhead{E($\bv$)}
&\colhead{M$_{wd}$ (M$_{\sun}$)}
&\colhead{M$_{2}$ (M$_{\sun}$)}
&\colhead{d (pc)}
}
\startdata
$^{1,2,3}$BK Lyn   & 14.6 & 15.3   & 0.075    & 32$\pm$12 & \nodata       & 0.3:          & \nodata       & 116-291 \\ 
$^{4,5,6}$V380 Oph & 14.3 & $> 17$ & 0.154107 & 42$\pm$13 & \nodata       & 0.58$\pm$0.19 & 0.36$\pm$0.04 & 293-513 \\
$^{a,b}$V751 Cyg   & 13.2 & 17.8   & 0.144584 & 30$\pm$20 & 0.25$\pm$0.05 & \nodata       & \nodata       & 201-352 \\ 
\enddata
\tablerefs{\\1 \citet{Dobrzycka1992}; 2 \citet{Ringwald1996}; 
3 \citet{Skillman1993} \\4 \citet{Shafter1983}; 
5 \citet{Shugarov2005}; 6 
\citet{Rodriguez-Gil2007}\\
a \citet{Echevarria2002} \\
b \citet{Greiner1999}}
\end{deluxetable}

\clearpage 

\begin{deluxetable}{lccccccc}
\tablecaption{{\it{HST}} Observations}
\tablewidth{0pc}
%\rotate
\tablenum{2}
\tablecolumns{7}
\tablehead{
\colhead{Target}              &
\colhead{Data ID}             &
\colhead{Config/Mode}            &
\colhead{Grating} &
\colhead{Obs. Date} &
\colhead{Obs. Time}  &
\colhead{t$_{exp}$(s)}
}
\startdata
BK Lyn   & O6LI34010 & STIS/FUV-MAMA/ACCUM & G140L & 2003-04-13 & 01:12:43 & 600 \\
V380 Oph & O6L186010 & STIS/FUV-MAMA/ACCUM & G140L & 2003-07-06 & 05:53:14 & 730 \\
V751 Cyg & O6LI1D010 & STIS/FUV-MAMA/ACCUM & G140L & 2002-12-01 & 17:11:16 & 700 \\
\enddata
\end{deluxetable}

\clearpage 

\begin{deluxetable}{lcccc}
\tablecaption{BK Lyn, V751 Cyg, 
V380 Oph Line Identifications from STIS Spectra}
\tablewidth{0pc}
\tablenum{3}
\tablehead{
\colhead{Ion}       &
\colhead{$\lambda$$_{rest}$}        &
\colhead{BK Lyn - $\lambda$$_{obs}$} &
\colhead{V751 Cyg - $\lambda$$_{obs}$} &
\colhead{V380 Oph - $\lambda$$_{obs}$} 
}
\startdata                         
C III &   1175.71  & 1175.39e & 1175.70a &  1175.00e   \\
N V &     1238.82  & 1238.57e & 1239.29a &  1237.14a   \\
          1242.80  & \nodata  & \nodata  &  1242a      \\     
Si II &   1260.42  & 1260.18e & 1260.87a &  1260.35a   \\
O I &     1302.17  & 1302.11a & 1302.28a &  1303.51a   \\ 
Fe V &    1329.69  & 1330.52e  & 1330.40a &  \nodata    \\
C II  &   1334.53  & 1334.94e & \nodata  &  1334.41a   \\   
C II &    1335.71  & \nodata  & 1334.94a &  \nodata    \\
O v &     1371.29  & 1371.49e  & 1372.25a &  1335.7077a \\
Fe v &    1376.34  & 1376.16e & 1375.75a &  \nodata    \\
Si IV &   1393.76  & \nodata & 1393.74a &  1395.48e   \\  
Si IV &   1402.77  & \nodata  & \nodata  &  1403.5e    \\
Fe V &    1479.17  & \nodata  & 1477.17a &  \nodata    \\
C IV &    1548.20  & 1548.26e & 1544.75a &  1548e      \\
     &    1550     & \nodata  & \nodata  &  1550.67e   \\
C I &     1560.68  & \nodata  & 1560.48a &  \nodata    \\
He II &   1640.47  & 1639.86e & 1645.61a &  1641.41e   \\
C I &     1657.01  & \nodata  & 1657.16a &  \nodata    \\      
Si I &    1696.16  & 1696.29e  & 1696.20a &  \nodata   
\enddata
\tablerefs{
\\a - denotes an absorption feature\\
e- denotes an emission feature\\}
\end{deluxetable}

\clearpage

\begin{deluxetable}{lcccc}
\tablecaption{Accretion Disk Model Fitting to the STIS Spectra of BK Lyn}
\tablenum{4}
\tablewidth{0pc}
\tablehead{
\colhead{ $M_{wd}~~~(M_{\odot})$ }                  & 
\colhead{ $\dot{M}_{model}~~(M_{\sun}$yr$^{-1}$)}   & 
\colhead{ $i~~(^\circ)$}                            & 
\colhead{ $\chi^2$}                                 & 
\colhead{d(pc)}
}
\startdata
1.2    &   -9/-9.5 &  75-81  &  1.84-1.9  &    300-550 \\
1.0    &   -9.0    &  75-81  &  1.94      &    400-700  \\
0.8    &   -8.5    &  75-81  &  1.97      &    560-940 \\
0.55   &   -8.0    &  75-81  &  1.97-2.00 &    730-1220 \\
0.40   &   -8.0    &  75     &  2.07      &    860 \\
\enddata
\end{deluxetable}

%\section{Figure Captions}

\clearpage

\begin{figure}
\vspace{-15.cm} 
\plotone{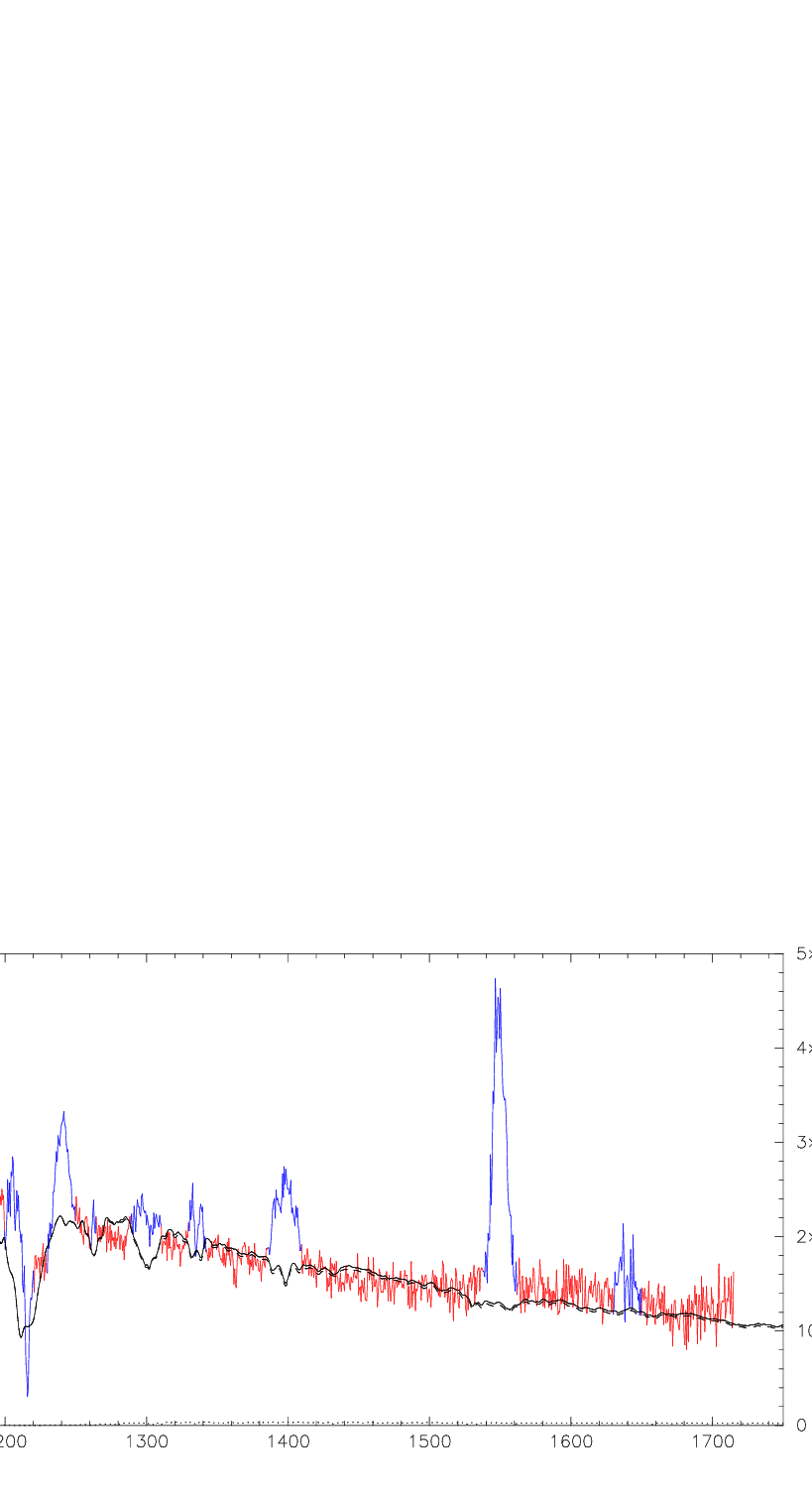}
\caption{ The best fitting accretion disk model fit to the HST STIS spectrum
of BK Lyn. This fit yields $\dot{M}=1 \times 10^{-9.0}$ with $d = 300$ pc, 
$M_{wd} = 1.2 M_{\sun}$, and $i=75^{\circ}$.} 
\end{figure}

\clearpage

\begin{figure}
\vspace{-15.cm} 
\plotone{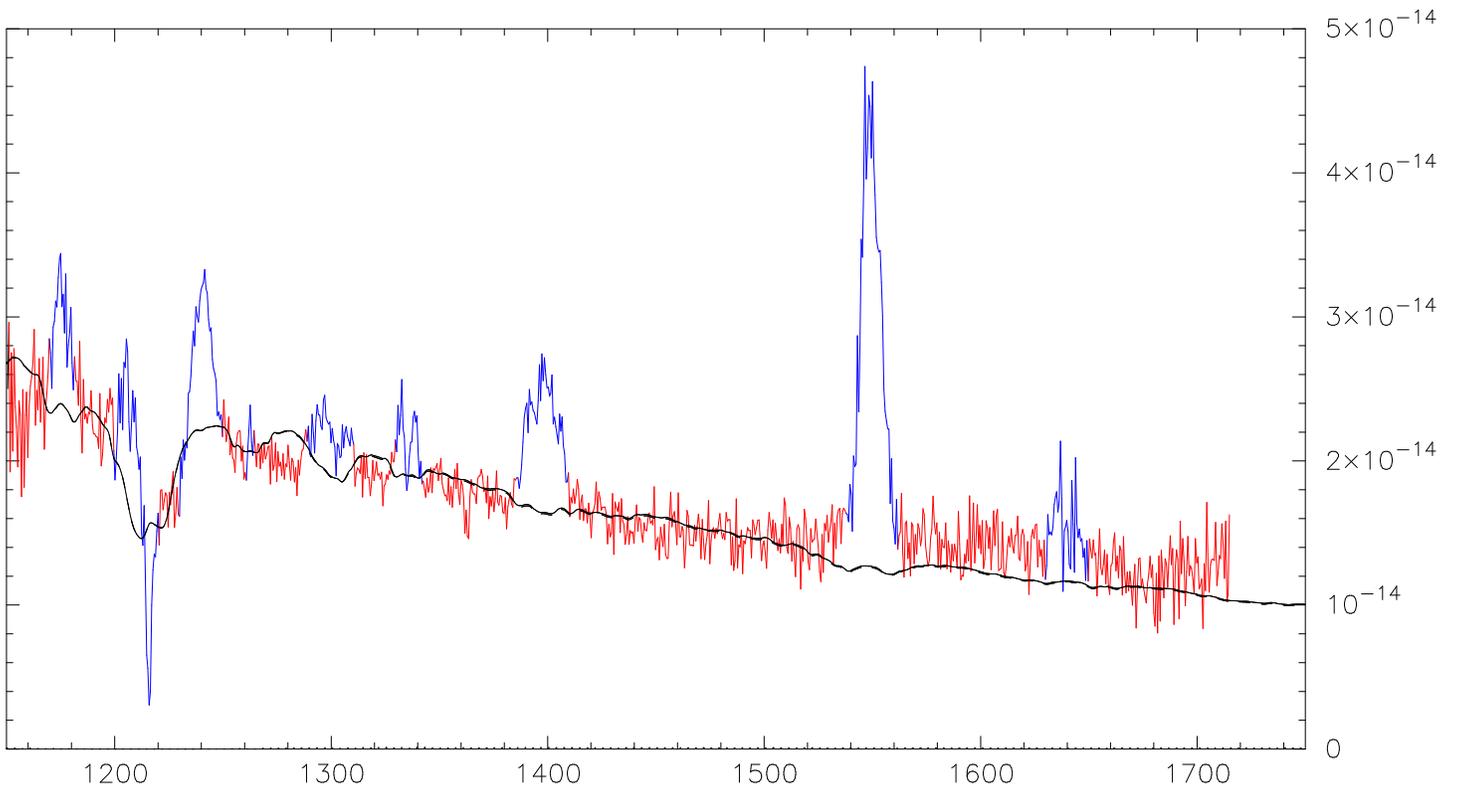} 
\caption{ The best fitting accretion disk model fit to the HST STIS spectrum
 of BK Lyn. This fit yields $\dot{M}$ = 1 $\times 10^{-8.5}$ with 
$d$ = 300 pc, M$_{wd}$ = 0.8 M$_{\sun}$. }
\end{figure}

\clearpage

\begin{figure}
\vspace{-15.cm} 
\plotone{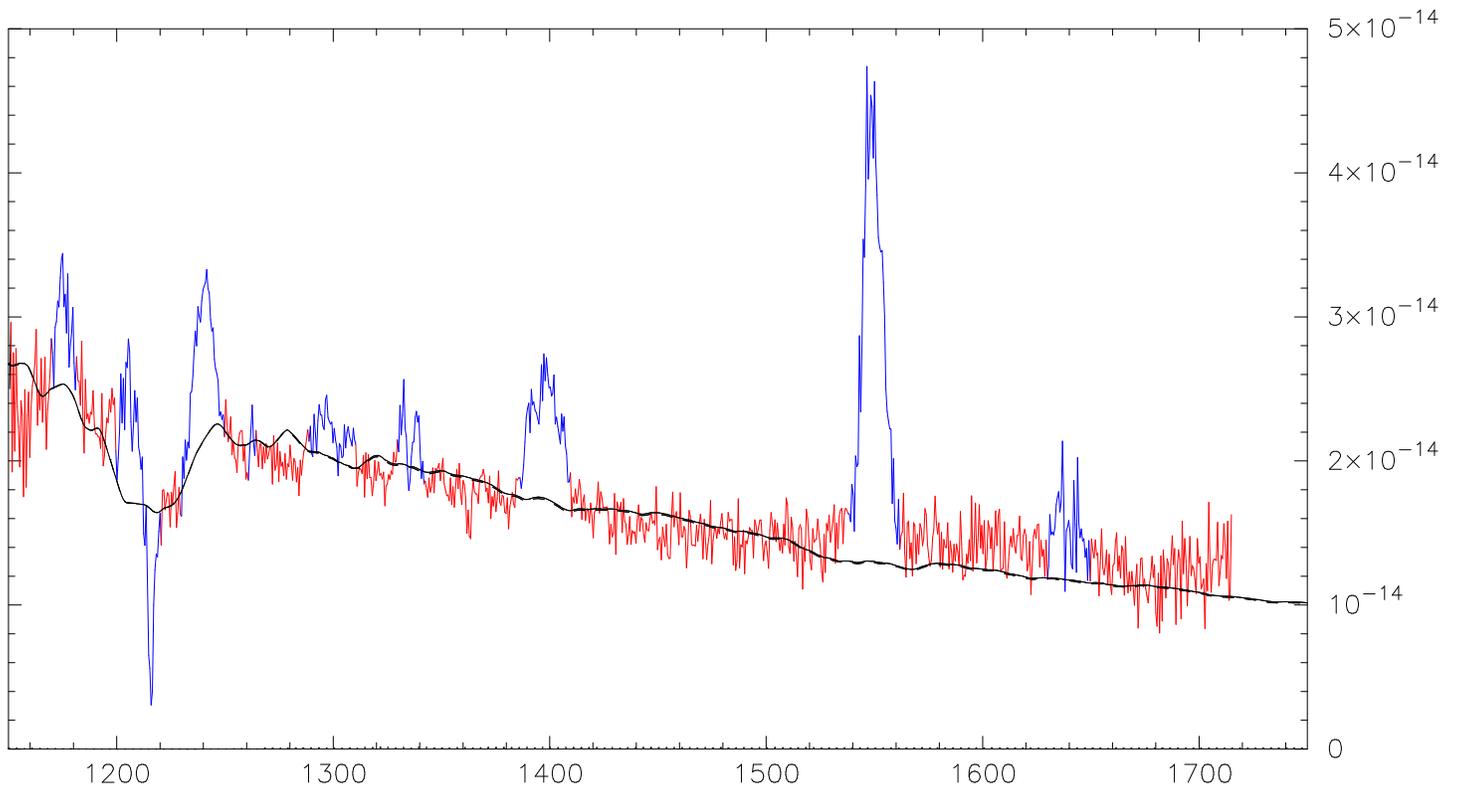} 
\caption{ The best fitting accretion disk model fit to the HST STIS spectrum
 of BK Lyn. This fit yields the parameters $d$ = 2046 pc, M$_{wd}$ = 
0.40 M$_{\sun}$, $i$ = 75$\degr$, 
and $\dot{M}$ = 1 $\times 10^{-8}$.}
\end{figure}

\clearpage

\begin{figure}
\vspace{-15.cm} 
%\epsscale{.8}
\plotone{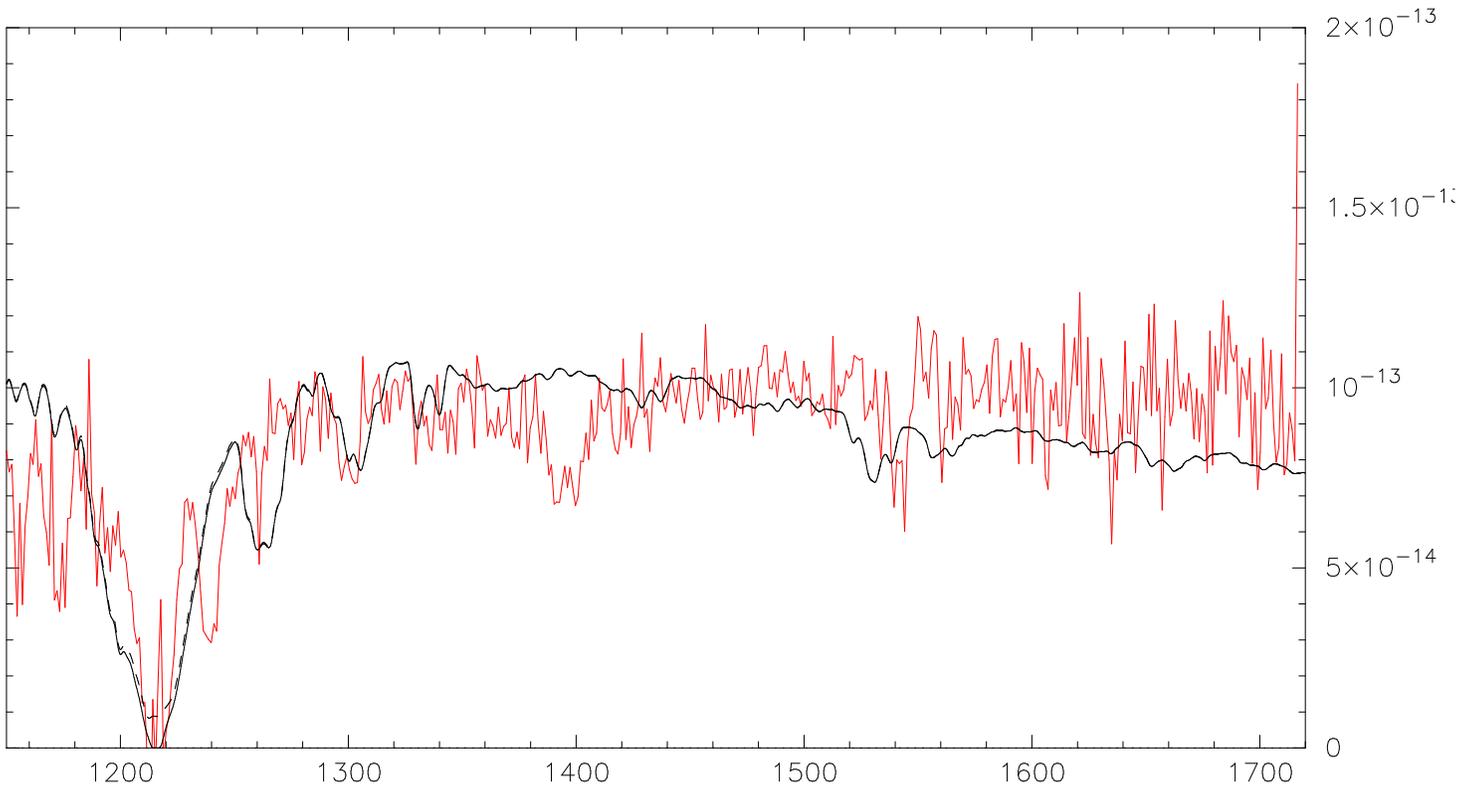} 
\caption{Comparison of a standard accretion disk model with the 
dereddened HST STIS
spectrum of V751 Cygni during its high optical brightness state 
illustrating the extent of disagreement.
The disk model has $M_{wd}=0.55M_{\odot}$, $i=75^{\circ}$, 
$\dot{M} = 1 \times 10^{-9}M_{\odot}$yr$^{-1}$ but a distance of
only 78pc!
}
\end{figure}

\clearpage

\begin{figure}
\vspace{-15.cm} 
%\epsscale{.8}
\plotone{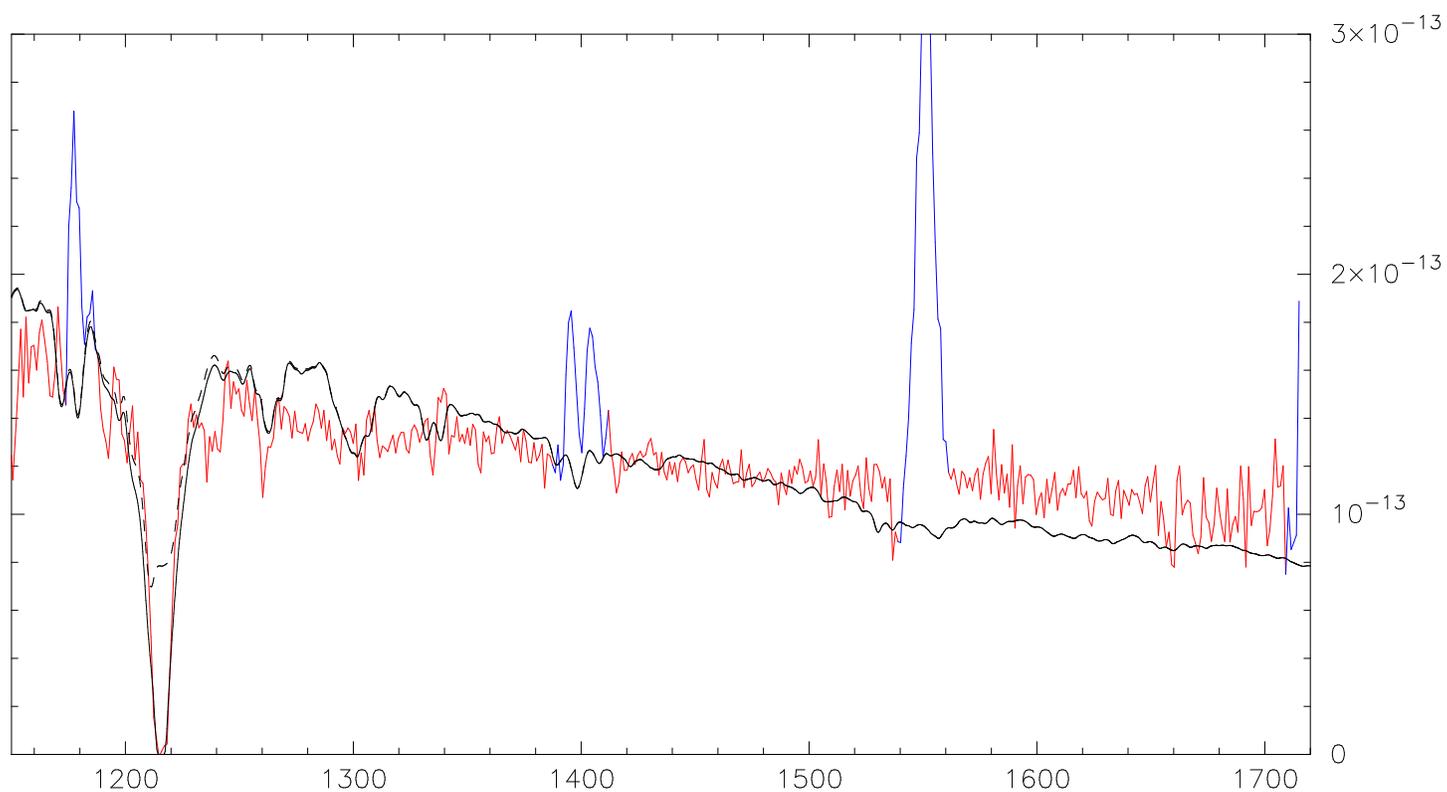} 
\caption{Comparison of a standard accretion disk model with the 
dereddend HST STIS
spectrum of V380 Oph during its high optical brightness state illustrating 
the extent of disagreement. 
The disk model has $M_{wd}=0.40M_{\odot}$, $i=75^{\circ}$, 
$\dot{M} = 1 \times 10^{-8}M_{\odot}$yr$^{-1}$ and a distance of
312pc. The fit to the Ly$\alpha$ profile has been
achieved with the inclusion of an ISM model (solid line). The disk model
alone, without the ISM inclusion does not reach the bottom of the
Ly$\alpha$ feature (dashed-line).}
\end{figure}

\clearpage

\begin{figure}
\vspace{-15.cm} 
\plotone{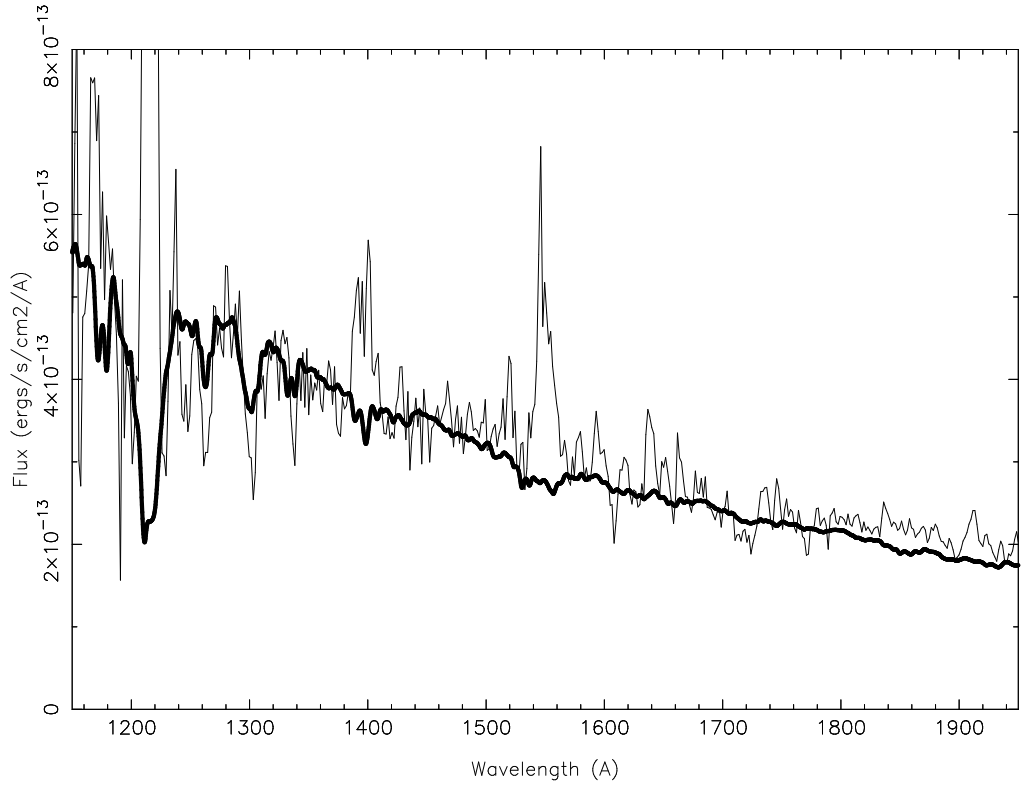} 
\caption{ 
The IUE spectra of the SW Sex star V442 Oph with an accretion 
disk model, from \citet{Ballouz2009}. The fit suggests that 
optically thick steady state disk models do provide a fair representation
of the FUV energy distribution and Ly$\alpha$ profile for some 
SW Sex stars.}
\end{figure}

\clearpage

\begin{figure}
\vspace{-15.cm} 
\plotone{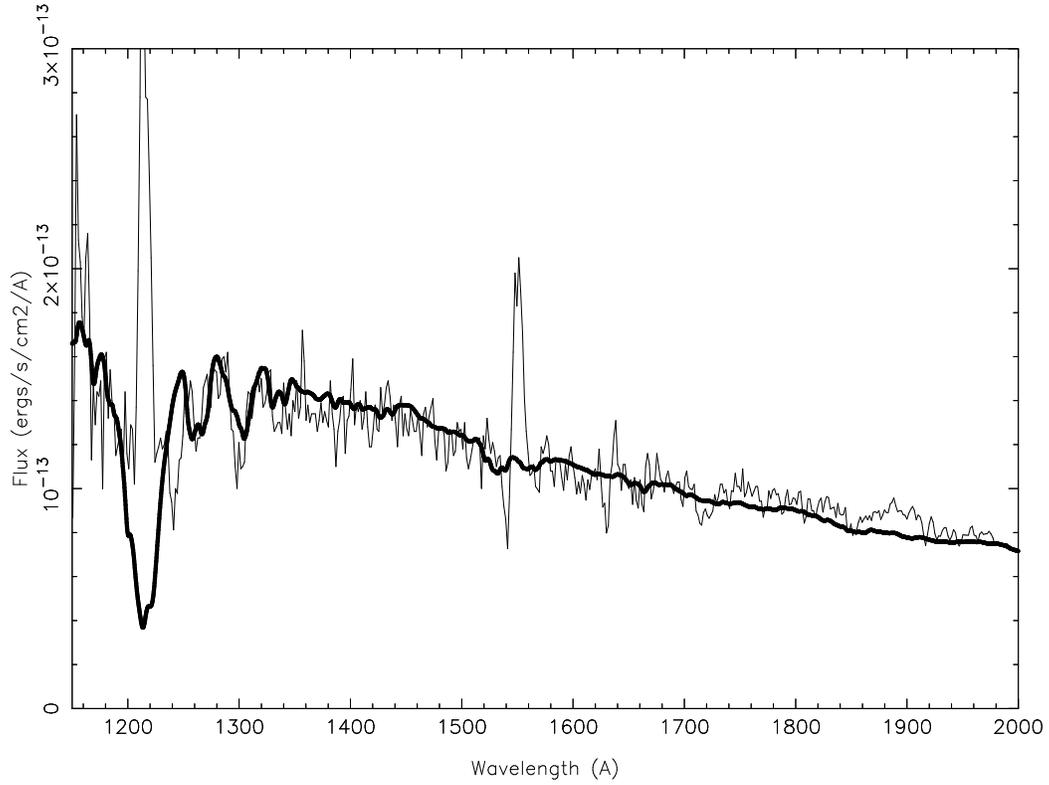} 
\caption{ 
The IUE spectrum of the SW Sex star AH Men with an accretion 
disk model, from \citet{Ballouz2009}. The fit demonstrates that 
optically thick steady state disk models give reasonable agreement
with the FUV energy distribution and Ly$\alpha$ profile of some 
SW Sex stars. 
}
\end{figure}


\begin{thebibliography}{}

\bibitem[Araujo-Betancor et al. (2005)]{Araujo-Betancor2005}
Araujo-Betancor,S., G\"ansicke, B.T., Long, K. S., Beuermann, K., 
de Martino, D., Sion, E. M., Szkody, P. 2005 \apj, 622, 589

\bibitem[Ballouz \& Sion (2009)]{Ballouz2009}
Ballouz, R.L, \& Sion, E.M. 2009, ApJ, in press 

\bibitem[Bruch \& Engel (1994)]{Bruch1994}
Bruch, A., \& Engel, A. 1994, \aaps, 104, 79 

\bibitem[Dhillon et al. (2000)]{Dhillon2000}
Dhillon, V.S., Littlefair, S.P., Howell, S.B., Ciardi, D.R., 
Harrop-Allin, M.K., \& Marsh, T.R. 2000, \mnras, 314, 826

\bibitem[Diaz et al. (1996)]{Diaz1996}
Diaz, M.P., Wade, R., Hubeny, I. 1996, \apj, 459, 236

\bibitem[Dobrzycka \& Howell (1992)]{Dobrzycka1992}
Dobrzycka, D., \& Howell, S.B. 1992, \apj, 388, 614 

\bibitem[Downes et al. (1995)]{Downes1995}
Downes, R., Hoard, D.W., Szkody, P., \& Wachter, S. 1995, \aj, 110, 1824

\bibitem[Echevarr\'ia et al. (2002)]{Echevarria2002}
Echevarr\'ia, J., Costero, R., Tovmassian, G., Zharikov,
S., Pineda, L., Michel, R, Rev.Mex.A.A. (Serie de Conferencias), 2002, 12, 
eds. W. Henney, J. Franco, M. Martos, \& M. Pe$\bar{n}$a, pp. 86-89

\bibitem[Godon et al.(2007)] {Godon2007}
Godon, P., Sion, E.M., Barrett, P.E., Szkody, P. 2007, ApJ, 656, 1029 

\bibitem[Greiner (1998)]{Greiner1998}
Greiner, J. 1998, \aap, 336, 626

\bibitem[Greiner et al. (1999)]{Greiner1999}
Greiner, J., Tovmassian, G, DiStefano, R., Prestwich,
A., Gonzalez-Riestra, R., Szentasko, L., Chavarria, C.
1999, \aap, 343, 183

\bibitem[Hamilton \& Sion (2008)]{Hamilton2008}
Hamilton, R.T., \& Sion, E.M. 2008, \pasp, 120, 165 

\bibitem[Hoard et al. (2003)]{Hoard2003}
Hoard et al.2003, \aj, 126, 2473

\bibitem[Hubeny (1988)]{Hubeny1988}
Hubeny, I. 1988, Comput. Phys. Comm., 52, 103

\bibitem[Hubeny (1990)]{Hubeny1990}
Hubeny, I. 1990, \apj, 351, 632

\bibitem[Hubeny \& Lanz (1995)]{Hubeny1995}
Hubeny, I., \& Lanz, T. 1995, \apj, 439, 875

\bibitem[Knigge (2006)]{Knigge2006}
Knigge, C. 2006, \mnras, 373, 484

\bibitem[La Dous (1991)]{LaDous1991}
La Dous, C. 1991, \aap, 252, 100

\bibitem[Livio \& Pringle (1994)]{Livio1994}
Livio, M., Pringle, J. E. 1994, \apj, 427, 956

\bibitem[Patterson et al. (2001)]{Patterson2001}
Patterson, J., Thorstensen,J., Fried,R.,  Skillman D., Cook,K. Jensen, K.
2001 \pasp, 113, 779, 72

\bibitem[Puebla et al. (2007)]{Puebla2007}
Puebla, R.E., Diaz, M.P., \& Hubeny, I. 2007, \aj, 134, 1923

\bibitem[Ringwald et al. (1996)]{Ringwald1996}
Ringwald, F.A., Thorstensen, J.R., Honeycutt, R.K., \& Robertson, J.W.
1996, \mnras, 278, 125 

%\bibitem[]{}
%Ritter, H., \& Kolb, U. 2003, A\&A, 404, 301

\bibitem[Rodriguez-Gil et al. (2007)]{Rodriguez-Gil2007}
Rodriguez-Gil, P., Schmidtobreick, L., G\"ansicke, B.T. 2007, 
\mnras 374, 1359

\bibitem[Shafter (1983)]{Shafter1983}
Shafter, A.W. 1983, Ph.D. thesis, UCLA

\bibitem[Shugarov et al. (2005)]{Shugarov2005}
Shugarov, S. Yu., Katysheva, N.A., Seregina, T.M., Volkov, I.M., Kroll, P. 
2005, in: The Astrophysics of Cataclysmic Variables and Related 
Objects, ASP Conference Vol. 330, Edited by J.-M. Hameury and J.-P. 
Lasota. San Francisco: Astronomical Society of the Pacific, 
2005., p.495 

\bibitem[Sion et al. (1995)]{Sion1995}
Sion, E. M., Huang, M., Szkody, P., Cheng, F. 1995 \apj, 444, 97

\bibitem[Skillman \& Patterson (1993)]{Skillman1993}
Skillman, D.R., \& Patterson, J. 1993, \apj, 417, 298 

\bibitem[Sproats et al. (1996)]{Sproats1996}
Sproats, L.N., Howell, S.B., \& Mason, K.O. 1996, \mnras, 282, 1211

\bibitem[Verbunt (1987)]{Verbunt1987}
Verbunt, F. 1987, \aaps, 71, 339 

\bibitem[Wade \& Hubeny (1998)]{Wade1998}
Wade, R.A., \& Hubeny, I. 1998, \apj, 509, 350 

\bibitem[Walker \& Bell (1980)]{Walker1980}
Walker, M., \& Bell, M. 1980, BAAS, 12, 63

\bibitem[Warner (1995)]{Warner1995}
Warner, B. 1995, Cataclysmic Variable Stars (Cambridge University Press)

%\bibitem[]{}
%Warner, B. 2008, private communication

\bibitem[Wood (1990)]{Wood1990}
Wood, M.A. 1990, Ph.D. thesis, University of Texas at Austin 

\bibitem[Wu et al. (1995)]{wu1995}
Wu, K., Wickramasinghe, D., \& Warner, B. 1995, PASA, 12, 604

\end{thebibliography}
\end{document}